\documentclass[12pt,preprint]{aastex}

\lefthead{Smith et al.}
\righthead{Fluorine Abundances in M4}

\received{2005 June 9}
\begin{document}

\title{Fluorine Abundance Variations in Red Giants of the Globular Cluster M4
and Early-Cluster Chemical Pollution}

\author{Verne V. Smith}
\affil{National Optical Astronomy Observatory, P.O. Box 26732, Tucson,
AZ 85726, USA; vsmith@noao.edu}

\author{Katia Cunha}
\affil{Observat\'orio Nacional, Rua General Jos\'e Cristino 77, 20921-400,
S\~ao Crist\'ov\~ao, Rio de Janeiro, Brazil; katia@on.br}

\author{Inese I. Ivans\altaffilmark{1}}
\affil{Department of Astronomy, California Institute of Technology, 1200
E. California Blvd., MS 105-24, Pasadena, CA 91125; iii@astro.caltech.edu}

\author{John C. Lattanzio}
\affil{Centre for Stellar \& Planetary Astrophysics, School of Mathematical
Sciences, PO Box 28M, Monash University, Victoria 3800, Australia;
John.Lattanzio@sci.monash.edu.au}

\author{Kenneth H. Hinkle}
\affil{National Optical Astronomy Observatory, P.O. Box 26732, Tucson,
AZ 85726, USA; khinkle@noao.edu}

\altaffiltext{1}{
Hubble Fellow
}

\begin{abstract}

We present fluorine abundances in seven red-giant members of the globular
cluster M4 (NGC 6121). These abundances were derived from the HF (1--0) R9
line at 2.3357$\mu$m in high-resolution infrared spectra obtained with
the Phoenix spectrograph on Gemini-South. Many abundances in the target stars
have been studied previously, so that their overall
abundance distributions are well-mapped. The abundance of fluorine is
found to vary by more than a factor of 6, with the $^{19}$F variations
being correlated with the already established oxygen variations, and
anti-correlated with the sodium and aluminium variations.
In this paper we thus add fluorine to the list of elements known to vary in
globular cluster stars, and
this provides further evidence
that H-burning is the root cause of the chemical inhomegeneities.
The fact that $^{19}$F is found to decrease in the M4 stars, as the
signature of H-burning appears,
indicates that the polluting stars must have
masses greater than about 3.5M$_{\odot}$, as less massive stars than
this should produce, not destroy, fluorine.

\end{abstract}


\clearpage

\section{Introduction}

Many galactic globular clusters display abundance patterns
not observed among field stars possessing similar masses
and metallicities. In a given cluster, star to star abundance
variations may be observed among a specific, small set of
light elements: C, N, O, Na, Al and sometimes Mg. Within
a given cluster, the C and N abundances are anti-correlated,
while O and Na, as well as Al and Mg,
are also anti-correlated. Such anti-correlations can be understood
to result from H-burning (or proton captures) among the various
reaction networks known as the CNO cycles, the Ne-Na cycle, and the
Mg-Al cycle. With each progressive set of cycles among increasingly
charged nuclei, the H-burning samples ever hotter regions
within a star. An extensive review of these abundance patterns,
along with their corresponding large set of literature, can be
found in Gratton, Sneden, \& Carretta (2004).

Because red giants are bright, mapping the globular
cluster abundance patterns has focused historically on giants.
In addition, red giants have convective envelopes that mix
material to their surfaces that has undergone some
H-burning; specifically, anti-correlations between C and N
are predicted by theory and observed spectroscopically
(e.g. Sweigert \& Mengel 1979; Charbonnel et al. 1998; Gratton et al. 2000).
Variations in O, Na, Mg,
or Al in low-mass
giants requires much deeper mixing, and such
deep mixing is not predicted by models of low-mass
stars. Nonetheless, extra-mixing scenarios can be developed
for low-mass red giants and these processes can be invoked to
explain the observed anomalous globular cluster abundance patterns (e.g.
Langer \& Hoffman 1995).

Invoking extra-mixing
does not explain why such processes occur in low-mass globular
cluster giants, yet are absent in field giants with similar masses,
metallicities, and luminosities. In addition, nagging observations
of anti-correlated C and N abundances in turn-off and main-sequence
globular cluster stars from low-resolution spectra
(e.g. Suntzeff \& Smith 1991), as well
as N and Na correlations in some turn-off stars (Briley et al.
1996) are not explained by in-situ red giant
deep mixing.  More recently, a number of high-resolution spectroscopic
studies have uncovered the same abundance variations in globular cluster
stars near the main sequence as those observed in the same cluster
red-giant stars (e.g., Gratton et al. 2001; Ramirez et al. 2001; Ramirez \&
Cohen 2003). 
The set of abundance variations among
CNO and Na-Mg-Al now observed in globular cluster turn-off stars
effectively rules out the presently observed generation of low-mass
red giants as the culprits responsible for the anomalous cluster
abundances. One must look to another site
to produce the hot H-burning abundance patterns
specific to the globular clusters.

Another astrophysical site that includes H-burning,
mixing, and mass loss which could possibly contaminate
an early globular cluster environment is provided by massive
asymptotic giant branch (AGB) stars. In massive AGB
stars with deep convective
envelopes, the envelope bases undergo hot H-burning
(hot bottom burning, or HBB); such burning can produce
the required CNO and Na-Mg-Al patterns seen in globular clusters.
Another signature of HBB would be the destruction of fluorine via
$^{19}$F(p,$\alpha$)$^{16}$O.
Although $^{19}$F is produced in lower-mass AGB
stars (Jorissen et al. 1992),
this production becomes efficient destruction in HBB above masses
of about 3.5M$_{\odot}$.

If a generation of massive AGB stars polluted
a young cluster environment (where star formation was still
ongoing) with slow winds from HBB envelopes,
the O, Na, Mg, and Al patterns
might be explained.
Such a scenario would predict significant $^{19}$F depletions.
A clean test of this picture can be carried out using
high-resolution IR spectroscopy of HF lines near 2.2$\mu$m:
such a test in M4 is the goal of this study. The globular
cluster M4 was selected, as it is nearby
(being perhaps the closest globular cluster) so its members are
relatively bright, and its abundance pattern (including N, O, Na, Mg,
and Al) was studied extensively by Ivans et al. (1999).

\section{Observations}

The target giants in M4 were a subset of those from Ivans et al. (1999),
who analyzed a total of 34 members. A sample of 7 red giants
was selected to span the observed range in oxygen and sodium abundances
as found by Ivans et al. (1999). We also selected the cooler stars,
as these would have stronger HF lines.

High-resolution infrared spectra
were obtained with the Gemini South
Telescope plus the Phoenix spectrograph (Hinkle et al. 2002) during
queue observing in May 2004.
The data are single order echelle spectra with a
resolution R=$\lambda$/$\Delta$$\lambda$=50,000, corresponding
to a resolution element of $\sim$ 4-pixels.
The observed wavelength is within the
K-band, near $\sim$ 23,400\AA, containing the molecular HF(1--0) R9 line
and numerous CO lines from the 2--0 and 3--1 vibration bands.

The M4 giants are bright in the near-infrared
so that total exposure times of 20 minutes were
needed to obtain signal-to-noise ratios in
excess of $\sim$300. The target stars were observed in different positions on the slit.
In addition, hot stars were also observed.
These hot star spectra were needed in order to remove
telluric lines from the stellar spectra.
All the spectra were reduced to one-dimension using
standard IRAF routines. More details on the observations and
reduction procedures adopted
are found in Smith et al. (2002).

A sample Phoenix spectrum is shown in Figure 1 for one of the M4 red giants.
Some of the prominent absorption lines are identified, with
numerous $^{12}$C$^{16}$O lines, along with the
HF 1-0 R9 line, as well as a couple of atomic lines from Fe I and
Na I.

\section{Stellar Parameters \& Analysis}

The Ivans et al. (1999) study was
based upon optical high-resolution
spectra.
The critical parameters needed for an abundance
analysis are effective temperature (T$_{\rm eff}$), surface
gravity (parameterized as log g), microturbulent velocity ($\xi$),
and overall model atmosphere metallicity. Ivans et al. based their
T$_{\rm eff}$-scale on line-depth ratios, as described by Gray (1994).
Surface gravities were fixed by enforcing ionization
equilibrium for Fe I/Fe II and Ti I/Ti II.
These gravities are in generally good agreement with
those predicted from model color-magnitude diagrams.
Finally, microturbulent velocities were
set from the Fe I lines which spanned a large range in equivalent
width. The stellar parameters are listed in Table 1, along with
the Fe I and Fe II abundances: all taken from Ivans et al. (1999).
The stellar designations are from Lee (1977).

The same model atmospheres as used by Ivans et al. (1999) were employed
here, which are taken from the MARCS grid (Gustafsson
et al. 1975); these models have been found to be quite adequate for stars
having the effective temperaturers and gravities spanned by the sample
here.

The new data added to M4 are the high-resolution IR spectra containing
the HF 1--0 R9 line, plus a well-defined sample of $^{12}$C$^{16}$O
lines (Figure 1). Equivalent widths were measured for 6 CO lines
and these values are shown in Table 2. The designations, excitation
potentials ($\chi$) and gf-values are taken from Goorvitch \& Chackerian (1994 a,b).
With the oxygen abundances from Ivans et al. (1999), these CO lines
were used to determine $^{12}$C abundances in these stars.

Fluorine abundances were determined from the HF (1-0) R9 line using
spectrum synthesis. The line list adopted is described in Smith et al. (2002)
and Cunha et al. (2003).
Samples of observed and synthetic spectra
for two of our targets are shown in Figure 2.
Due to the high signal-to-noise of the spectra, the $^{19}$F
abundance is well determined. The dependance of the derived
fluorine abundances on the stellar parameters was computed and
is summarized as follows: $\delta$ T$_{eff}$ of +100K leads to a
increase of +0.22 dex in the fluorine abundance, $\delta$ log g = +0.3
produces a $\delta$F = -0.11 dex, and a $\delta$ $\xi$ = +0.5 km s$^{-1}$
results in 0.00 dex change in fluorine.
The changes were computed using a baseline model for L1514 of T$_{eff}$=3875 K,
Log g = 0.35, with $\xi$=2.5 km s$^{-1}$.

Abundances are summarized in Table 3, with the $^{12}$C and $^{19}$F
values derived here, while those of $^{13}$C were re-derived from
the M4 data of Ivans et al. (1999), employing our newly derived
 $^{12}$C values. Also included in the table are
O, Na, Mg, and Al taken
from Ivans et al. (1999). As the $^{12}$C abundances were derived from
6 lines, the standard deviations are also given in Table 3, which
gives some measure of the overall quality of the spectra and the
analysis.

The newly derived $^{12}$C abundances here update those of
Suntzeff
\& Smith (1991) and which were employed in Ivans et al. (1999).
The $^{12}$C abundances presented by Suntzeff \&
Smith were derived from low-resolution 2.2$\mu$m spectra which were
unsuitable for determining microturbulent velocities, thus assumed
values of $\xi$ had to be used. The actual values of $\xi$ determined
from high-resolution optical spectra by Ivans et al. (1999) and from
the high-resolution IR spectra here (using the $^{12}$C$^{16}$O
lines) are considerably lower than those assumed by Suntzeff \&
Smith (1991); on average for the 7 red giants observed here, the
values of microturbulent velocity are 0.7 km s$^{-1}$ lower. This
increases the $^{12}$C abundances significantly. Even with this
increase, the effect of decreasing carbon-12 abundance with increasing
red-giant luminosity found by Suntzeff \& Smith (1991) is confirmed
for the stars observed. We illustrate this in Figure 3, where
the $^{12}$C abundance is plotted versus the red-giant bolometric
magnitude.
   
The fluorine abundances show much larger scatter than
the abundance uncertainties.
Ivans et al. (1999) found large scatter in the other
abundances of C, N, O, Na, Mg and Al, as discussed in the
introduction. Here we show that fluorine behaves in a similar manner
to these other elements. The addition of fluorine to the list of
elements found to vary in the globular cluters has implications for
the types of progenitor stars that produce these abundance
variations.

In addition to fluorine, the newly derived  $^{12}$C abundances can be
added to those of  $^{13}$C,  $^{14}$N, and  $^{16}$O to form the
total sum of C+N+O and this is shown in Table 3. This sum is constant
to high accuracy for all stars studied with a mean and standard deviation
of C+N+O=8.17$\pm$0.08: the scatter can be accounted for entirely
by analysis errors. The constancy of C+N+O limits strongly the addition
of any primary  $^{12}$C from  $^{4}$He-burning and third dredge-up
on the AGB.    
\section{Discussion}

In the introduction we discussed how globular clusters show specific
abundance variations centered on the elements C, N, O, Na, Mg, and
Al (with a current summary in Gratton et al. 2004). With the increasing
numbers of globular cluster main-sequence stellar abundance studies
demonstrating that these abundance variations must arise from
early chemical evolution within the globular clusters, we will use
the M4 chemical inhomogeneities (including fluorine) to define
quantitative chemical yields required from these progenitor stars.

Figure 4 summarizes the abundance variations observed in M4.
We choose to use Na as a comparison element and then plot
[X/Fe] vs [Na/Fe] for the elements O, F, Mg, Al, Ti and La.
We do not include C and N in this plot, because low-mass
red-giant mixing alters these abundances anyway, thus
obscuring any primordial signatures. 
The filled symbols in Figure 4 are the stars observed here for
HF, while the open symbols are all other stars from Ivans et al.
(1999); approximate error bars are plotted for stars observed
for this study. 
As known previously, oxygen decreases
as Na increases, while Al increases with Na. We show here for
the first time that fluorine decreases as Na increases, with
an even larger variation than oxygen. As expected, Ti does not
vary with Na. La does not vary either, which precludes any significant
contribution from the s-process. Except for one star from the
Ivans' sample, Mg does not seem to change significantly with Na.

Adding fluorine to the suite of elements whose abundances
vary in M4 provides another constraint on the underlying
nucleosynthesis pattern. We proceed to define this pattern
based on the simple model for chemical enrichment in M4.
The assumption is that the oxygen-high, sodium-low stars
define the initial chemical mix in M4. Some mass range within
this stellar generation
added ejecta to the M4 intracluster environment which bore
the signature of hydrogen burning in the CNO, Ne-Na, Mg-Al
cycles. This material did not originate in massive
stars as there is no evidence of significant variations
in elements that arise from SN II (such ejecta could have
escaped from the intra-cluster medium due to high velocity).
In order to understand the chemical history in M4, it is crucial to pin down
the mass range of the polluting stars.
In this model a new generation
(or generations) of stars formed from differing fractions
of this stellar ejecta thus producing the observed
abundance inhomogeneities. We used the distributions in
Figures 4 to define the initial abundances of oxygen,
fluorine, sodium, magnesium, and aluminium. Final
abundances, which would represent the extreme case of
pure ejecta, are defined by the most Na-rich part
of the distributions. These initial and final abundances
are tabulated in Table 4, as well as the final - initial
abundances expressed as a mass fraction.

The mass fractions shown in Table 4 represent the yields
that were produced by the polluting stars. These yields
are qualitatively what is expected from H-burning at
temperatures of about 50 MK. What is of special
importance is the rather large depletion of fluorine.
Based on models by Fenner et al. (2004) stars with masses
less than about 3.5 solar masses are net producers of
fluorine during AGB evolution. Our results rule out stars
of this mass or less as being significant contributors to
chemical evolution within M4. These observationally
derived yields can be used to constrain even more
tightly the mass range of the polluting stars. As a simple
illustration, we show in Figure 5 predicted mass yields
from HBB AGB model stars as a function of stellar mass for
masses of 3.5, 5.0, and 6.5 solar masses for the elements
of interest here. Superimposed on this we also show
the estimated yields from Table 4. The importance
of fluorine can be seen from this comparison
as its yield is a steep function of stellar mass
and covers both positive and negative values.
This is a simple, but promising comparison illustrating
reasonable qualitative agreement, but quantitative
agreement remains lacking. At the observed level
of the fluorine and oxygen depletions, the Na, Mg and Al
enhancements predicted by the models appear to be too
large. Of course, the yields shown in Figure 5
need to be convolved with an IMF; however, a simple
convolution shows that the lower mass stars tend to
dominate.

The estimation of the observed abundance yields plotted
in Figure 5 are based upon a very simple chemical
evolutionary model for M4.  Better agreement between
the stellar models and the observed estimated chemical
yields might be achieved with additions to the simple
model.  For example, it might be that some of the
original stars had even lower Na and Al abundances than
those estimated here.  Another possibility is that
AGB stars even more massive than 6.5M$_{\odot}$ may
provide a better comparison.  There might also be less
dilution of CNO-processed material in the real stars
when compared to the models.  Observations of more
globular clusters, especially those having different
metallicities, will be useful in future comparisons of
observed abundance trends with stellar model predictions. 

\section{Conclusions}

We present the first fluorine abundances derived in a
globular cluster that show variations which correlate with
the oxygen abundances and anti-correlate with the Na and
Al variations. Because the predicted fluorine yields are
a strong function of stellar mass, fluorine abundances
can be used as a significant constraint in defining
the nature of the progenitor stars responsible for driving
chemical evolution in globular clusters. Knowledge
of the mass range of these stars would provide strong
limits on the timescales for early star formation in a
proto-globular cluster environment. Our results are
in qualitative agreement with HBB AGB models, but strict
quantitative agreement remains elusive; however,
the combination of more abundances in other clusters,
particularly fluorine, coupled with further efforts in
stellar modelling may yield the desired agreement.

\section{Acknowledgements}

Based on observations obtained at the Gemini Observatory,
which is operated by the Association
of Universities for Research in Astronomy, Inc., under a cooperative
agreement with the NSF on behalf of the Gemini partnership: the National
Science Foundation (United States), the Particle Physics and Astronomy
Research Council (United Kingdom), the National Research Council (Canada),
CONICYT (Chile), the Australian Research Council (Australia),
CNPq (Brazil), and CONICRT (Argentina), as program GS-2004A-Q-20. This
paper uses data obtained with the Phoenix infrared spectrograph,
developed and operated by the National Optical Astronomy Observatory.
This work is also supported in part by the National Science Foundation through
AST03-07534 (VVS), NASA through NAG5-9213 (VVS), and AURA, Inc. through
GF-1006-00 (KC).
I.I. is pleased to acknowledge research support from NASA
through Hubble Fellowship grant HST-HF-01151.01-A from the
Space Telescope Science Inst., operated by AURA, under NASA
contract NAS5-26555.  We wish to thank the anonymous referee
for useful comments that improved this paper.

\clearpage

\begin{deluxetable}{ ccccccc }
\setcounter{table}{0}
\tablewidth{320pt}
\tablecaption{Stellar Parameters}

\tablehead{ Star &
\multicolumn{1}{c} {T$_{\rm eff}$(K)} &
\multicolumn{1}{c} {Log g (cm s$^{-2}$)} &
\multicolumn{1}{c} {$\xi$ (km s$^{-1}$)} &
\multicolumn{1}{c} {A(Fe I)$^{a}$} &
\multicolumn{1}{c} {A(Fe II)$^{a}$}
}
\startdata
 1411 & 3950 & +0.60 & 1.65 & 6.29 & 6.32 \\
 1514 & 3875 & +0.35 & 1.95 & 6.27 & 6.42 \\
 2307 & 4075 & +0.85 & 1.45 & 6.30 & 6.33 \\
 3209 & 3975 & +0.60 & 1.75 & 6.28 & 6.32 \\
 3413 & 4175 & +1.20 & 1.65 & 6.32 & 6.35 \\
 4611 & 3725 & +0.30 & 1.70 & 6.34 & 6.33 \\
 4613 & 3750 & +0.20 & 1.65 & 6.31 & 6.35 \\
\enddata
\tablecomments{(a): A(X)= log[n(X)/n(H)] + 12. The solar iron abundance
is A(Fe)$_{\odot}$= 7.45.
}
\end{deluxetable}

\clearpage

\begin{deluxetable}{ ccccccccccc }
\setcounter{table}{1}
\tablewidth{460pt}
\tablecaption{$^{12}$C$^{16}$O Lines and Equivalent Widths}

\tablehead{ $\lambda$(\AA) &
\multicolumn{1}{c} {$\chi$(eV)} &
\multicolumn{1}{c} {Log gf} &
\multicolumn{1}{c} {Line ID} &
\multicolumn{1}{c} {1411$^{a}$} &
\multicolumn{1}{c} {1514$^{a}$} &
\multicolumn{1}{c} {2307$^{a}$} &
\multicolumn{1}{c} {3209$^{a}$} &
\multicolumn{1}{c} {3413$^{a}$} &
\multicolumn{1}{c} {4611$^{a}$} &
\multicolumn{1}{c} {4613$^{a}$}
}
\startdata
23302.949 & 1.421 & -4.517 & 3-1 R70 & 159 & 248 & 160 & 221 & 176 & 238 & 250 \\
23303.959 & 0.485 & -4.996 & 3-1 R30 & 392 & 476 & 368 & 468 & 408 & 456 & 482 \\
23340.590 & 0.431 & -5.064 & 3-1 R26 & 382 & 515 & 360 & 460 & 388 & 474 & 498 \\
23367.117 & 0.005 & -6.338 & 2-0 R4 & 241 & 388 & 208 & 314 & 247 & 336 & 346 \\
23372.385 & 0.396 & -5.124 & 3-1 R23 & 350 & 484 & 336 & 434 & 348 & 457 & 474 \\
23373.398 & 1.657 & -4.456 & 3-1 R77 & 130 & 233 & 126 & 196 & 152 & 183 & 168 \\
\enddata
\tablecomments{(a): Equivalent width in m\AA.
}
\end{deluxetable}

\clearpage

\begin{deluxetable}{ cccccccccc }
\setcounter{table}{2}
\tablewidth{460pt}
\tablecaption{Abundances$^{a}$}

\tablehead{ Star &
\multicolumn{1}{c} {A($^{12}$C)} &
\multicolumn{1}{c} {A($^{19}$F)} &
\multicolumn{1}{c} {$^{12}$C/$^{13}$C} &
\multicolumn{1}{c} {A($^{14}$N)} &
\multicolumn{1}{c} {A(O)} &
\multicolumn{1}{c} {A(Na)} &
\multicolumn{1}{c} {A(Mg)} &
\multicolumn{1}{c} {A(Al)} &
\multicolumn{1}{c} {A(C+N+O)}
}
\startdata
 1411 & 7.08 $\pm$0.08 & 2.90 & 4.0 & 7.70 & 7.93 & 5.45 & 6.66 & 6.01 & 8.177 \\
 1514 & 7.28 $\pm$0.11 & 3.15 & 5.0 & 7.18 & 8.18 & 5.07 & 6.77 & 5.70 & 8.277 \\
 2307 & 7.24 $\pm$0.07 & 3.10 & 5.0 & 7.47 & 7.91 & 5.42 & 6.78 & 5.97 & 8.119 \\
 3209 & 7.34 $\pm$0.07 & 3.20 & 5.0 & 7.20 & 8.00 & 5.25 & 6.85 & 5.93 & 8.153 \\
 3413 & 7.45 $\pm$0.10 & 3.35 & 4.0 & 6.96 & 8.16 & 5.01 & 6.74 & 5.87 & 8.276 \\
 4611 & 7.17 $\pm$0.04 & 2.70 & 5.0 & 7.72 & 7.81 & 5.39 & 6.70 & 5.86 & 8.130 \\
 4613 & 7.09 $\pm$0.05 & 2.88 & 4.0 & 7.55 & 7.81 & 5.31 & 6.72 & 5.88 & 8.062 \\
 Sun & 8.39 & 4.55 & 89 & 7.78 & 8.66 & 6.17 & 7.53 & 6.37 & 8.884 \\
\enddata
\tablecomments{(a): A(X)= log[n(X)/n(H)] + 12. The $^{12}$C, $^{13}$C and $^{19}$F
abundances are from this study, while the other abundances are from
Ivans et al. (1999).
}
\end{deluxetable}

\clearpage

\begin{deluxetable}{ ccccc }
\setcounter{table}{3}
\tablewidth{420pt}
\tablecaption{Observationally Derived Chemical Yields for M4}
                                                                                            
\tablehead{ Element &
\multicolumn{1}{c} {A(x)$_{init}$} &
\multicolumn{1}{c} {A(x)$_{fin}$} &
\multicolumn{1}{c} {Log (A(x)$_{fin}$ - A(x)$_{init}$)}
}
\startdata
O & 7.98 & 7.62 & -0.36 \\
F & 3.35 & 2.68 & -0.67 \\
Na & 4.82 & 5.49 & +0.67 \\
Mg & 6.74 & 6.76 & +0.02 \\
Al & 5.66 & 5.95 & +0.29 \\
Ti & 4.00 & 4.01 & +0.01 \\
La & 0.38 & 0.38 & 0.00 \\
\enddata

\end{deluxetable}
                                                                                            
\clearpage
\begin{figure}
\epsscale{.80}
\plotone{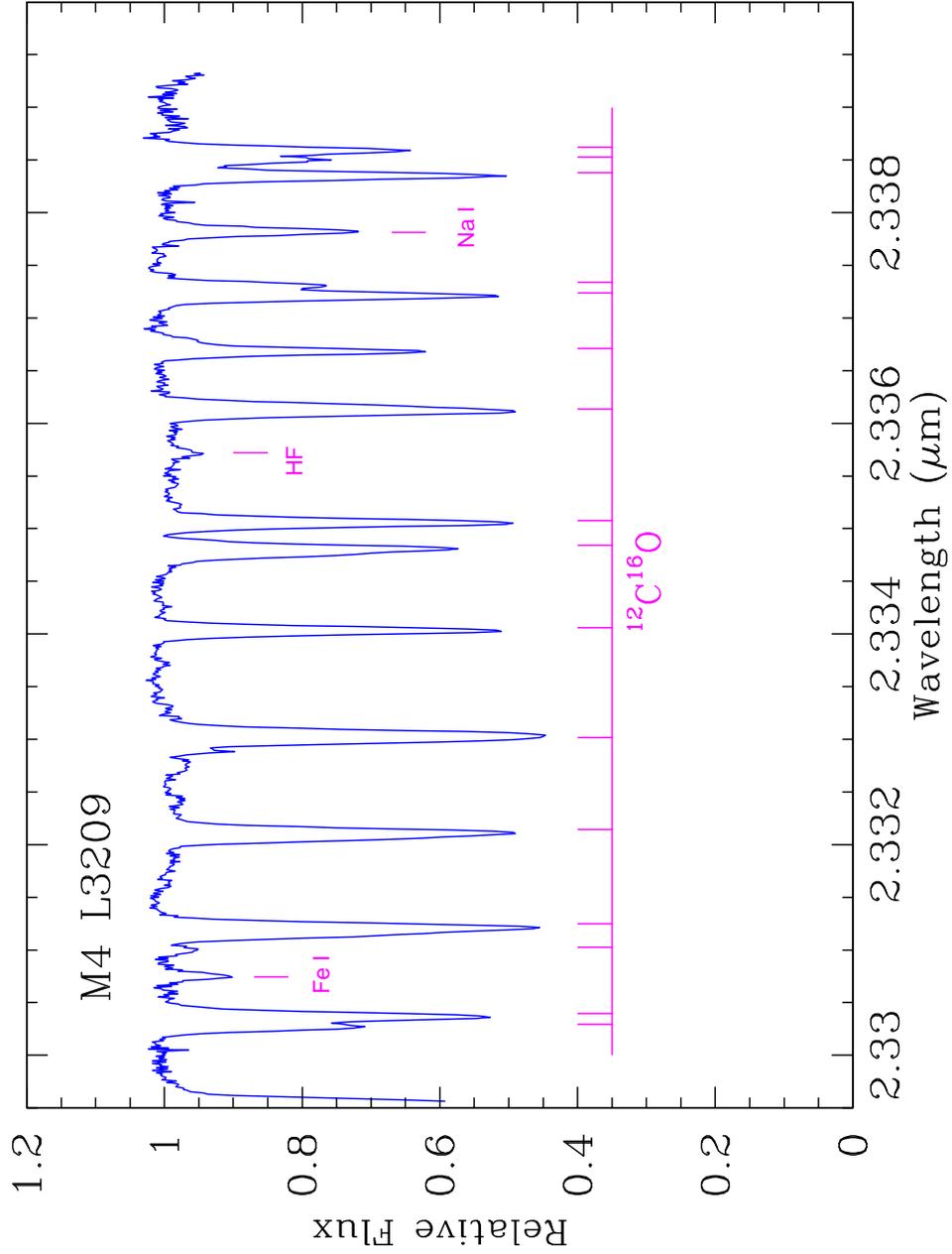}
\figcaption[fig1.ps]{Observed Phoenix spectra for the M4 red giant
Lee 1411 centered at 23390 \AA.
Prominent absorption lines are labelled.
\label{fig1}}
\end{figure}

\clearpage
\begin{figure}
\plotone{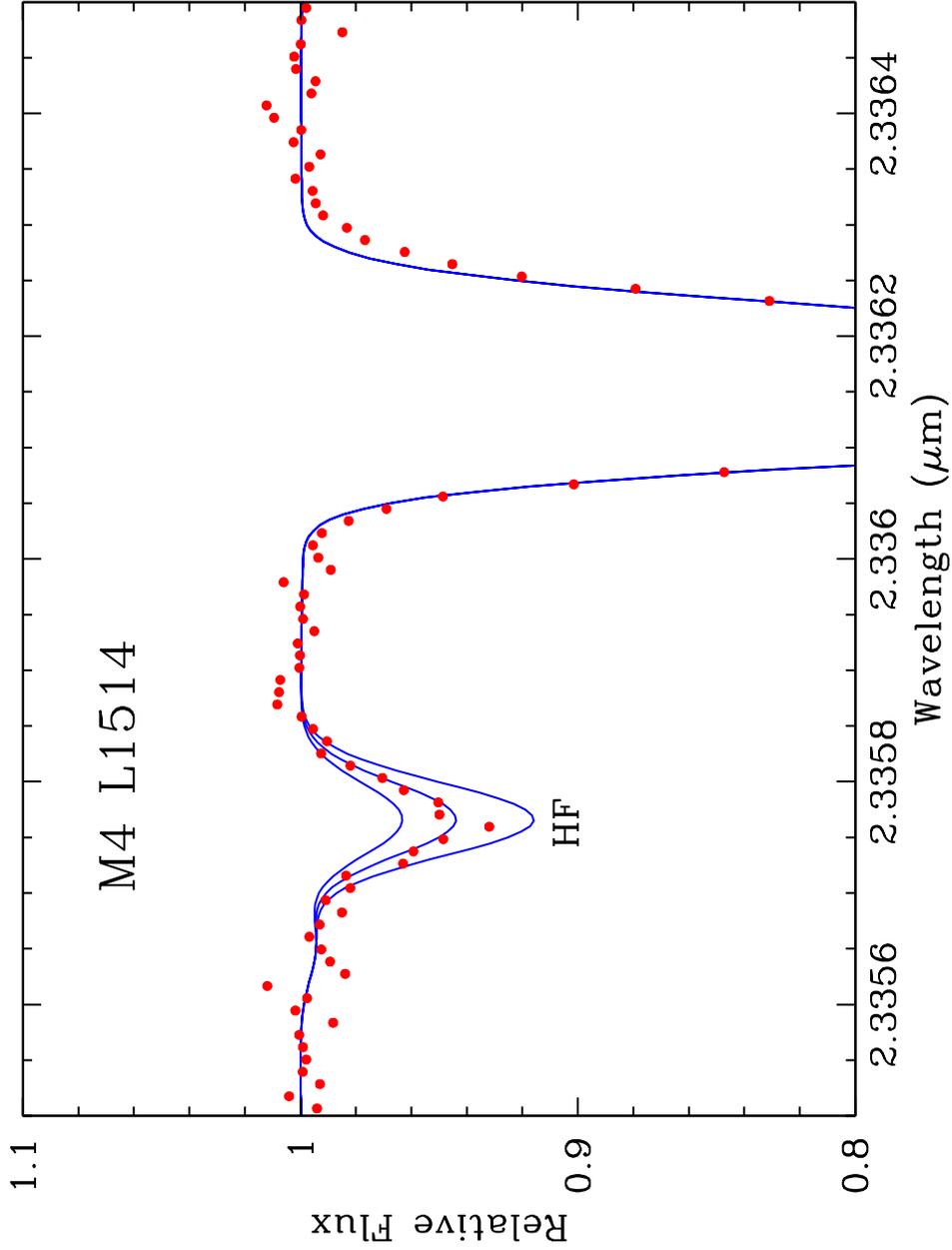}
\figcaption[fig2.ps]{Observed and synthetic spectra for target
star M4 Lee1514 in the HF region. The best-fit fluorine abundances (represented
by the middle solid line) ia that presented in Table 3. The synthetic
spectra were calculated with
$\pm$ 0.2dex changes in the $^{19}$F abundance.
\label{fig2}}
\end{figure}

\clearpage
\begin{figure}
\plotone{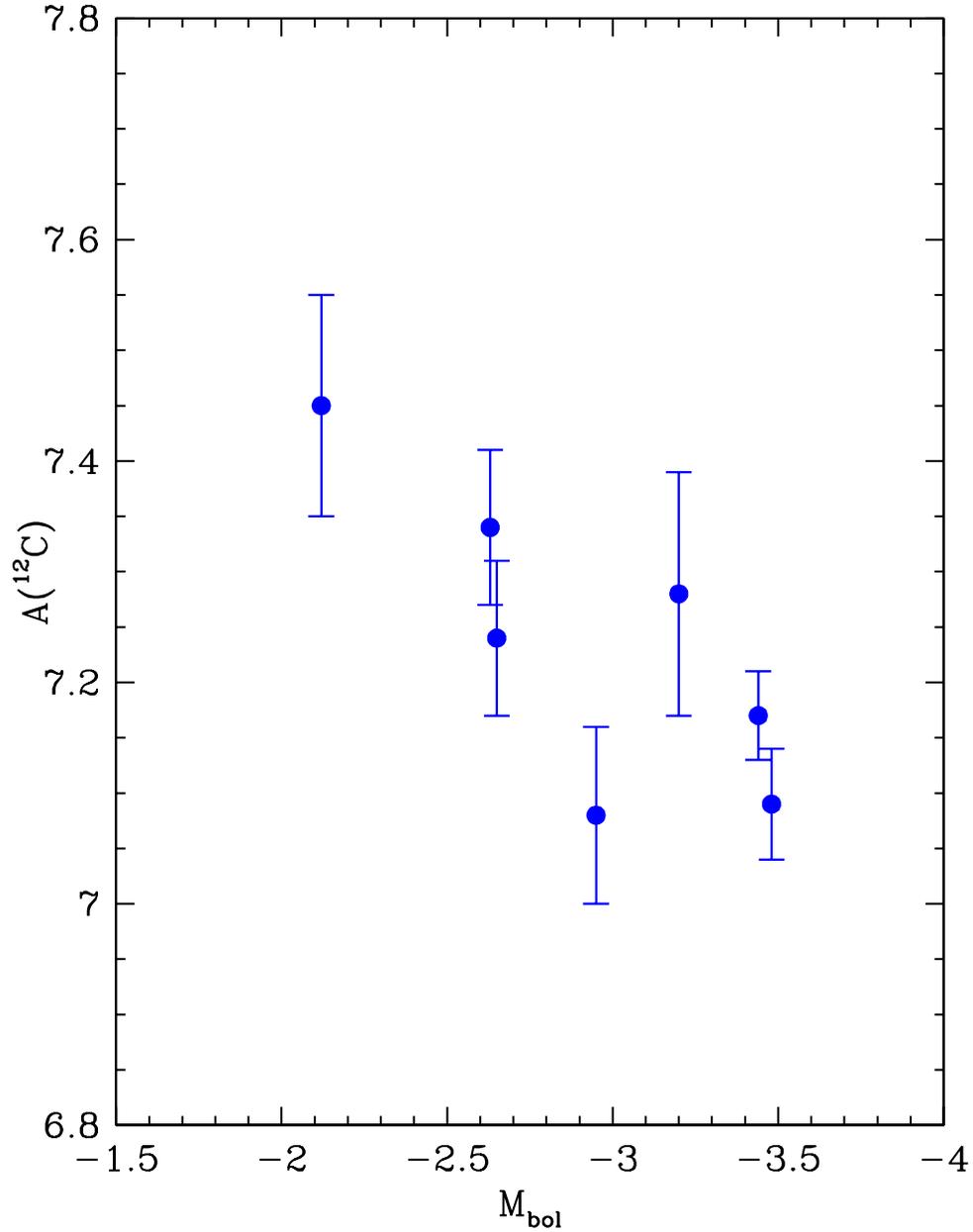}
\figcaption[fig3.ps]{Carbon-12 abundances versus M$_{bol}$ for the seven
M4 red giants observed here (M$_{bol}$ values are taken from Ivans et al.
1999). The decreasing $^{12}$C abundance with increasing red-giant
luminosity noted for M4 by Suntzeff \& Smith (1991) is also found using
updated $^{12}$C abundances.
\label{fig3}}
\end{figure}

\clearpage
                                                                               \begin{figure}                  
\plotone{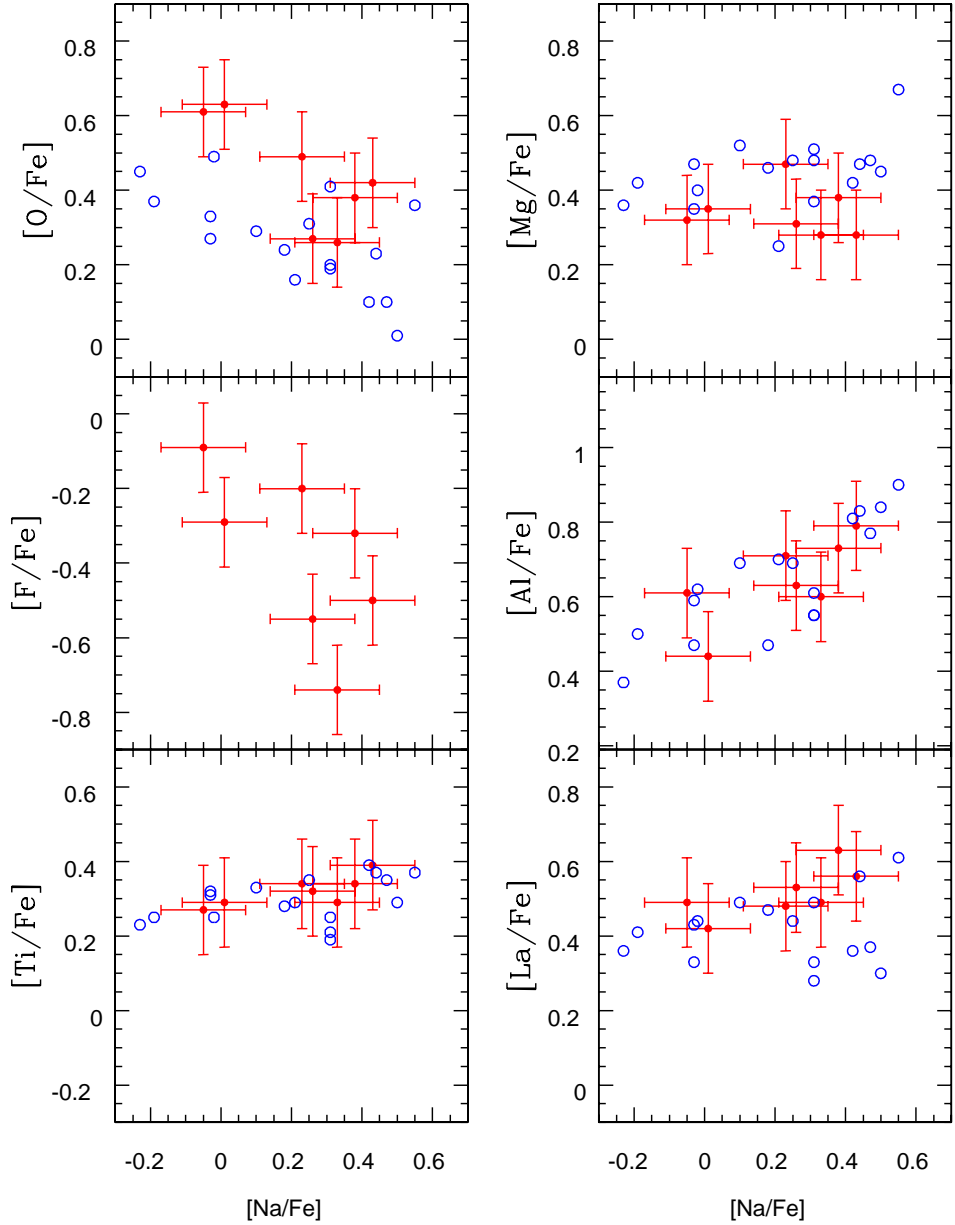}
\figcaption[fig4.ps]{Six elemental abundance ratios (measured relative to Fe)
plotted versus [Na/Fe] for stars in M4. The solid circles are target
stars observed for HF and the open circles are all other stars from Ivans et al.
(1999). Note the significant anti-correlation of [F/Fe] with [Na/Fe], as was
found previously for [O/Fe] versus [Na/Fe], while [Al/Fe] increases
with [Na/Fe]. No significant trends (of about  0.05 to 0.10 dex per dex)
are observed for [Mg/Fe], [Ti/Fe], or [La/Fe] with [Na/Fe].
\label{fig4}}
\end{figure}

\clearpage
                                                                               \begin{figure}                 
\plotone{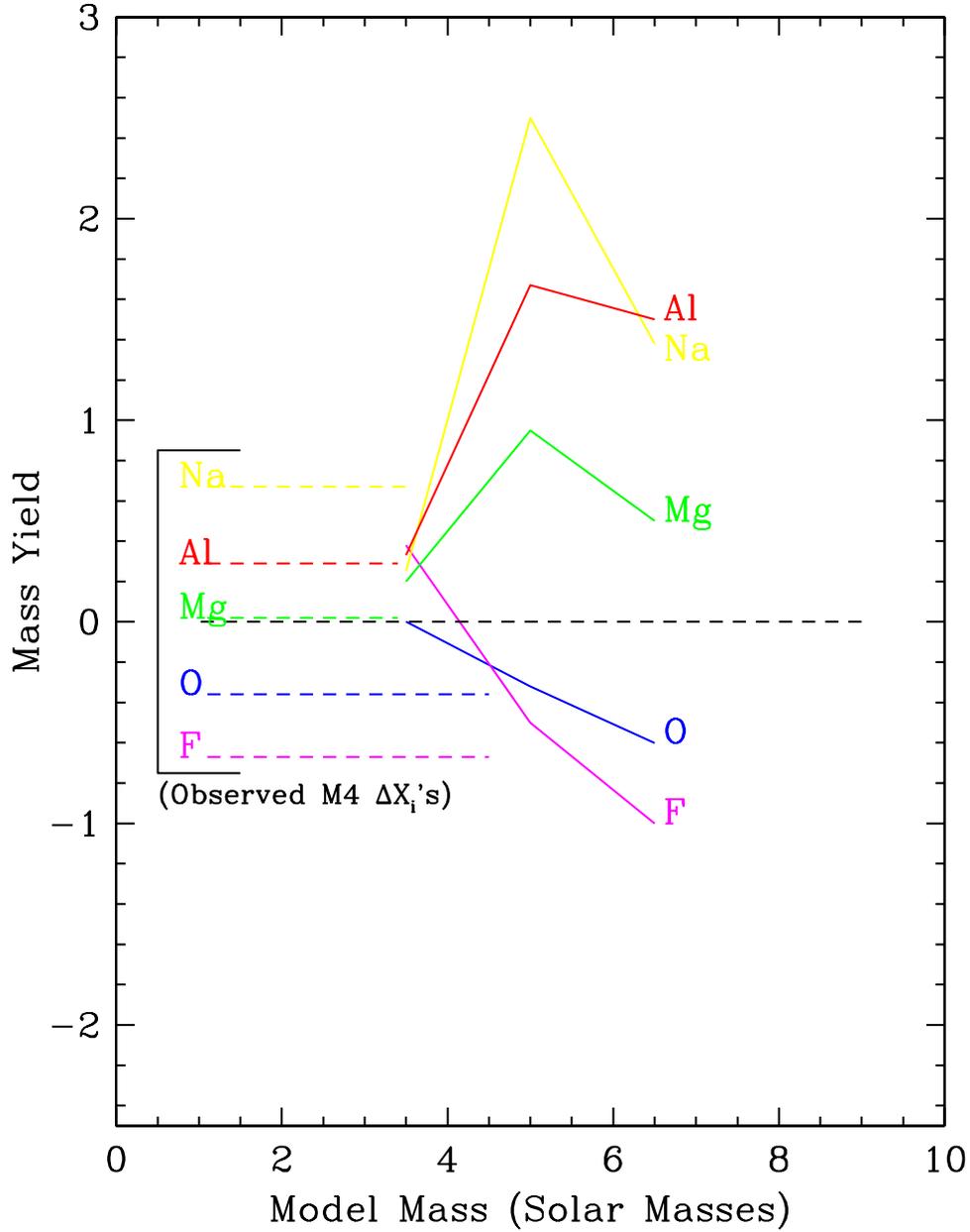}
\figcaption[fig5.ps]{Predicted mass yields from stellar models undergoing
HBB on the AGB (solid lines); these are taken from Fenner et al. (2004)
which were used in an analysis of NGC6752 abundances. The dashed lines are
the observationally derived mass yields for M4 from this study. Detailed
agreement between observations and models is still lacking, however, there
is rough agreement for the bulk abundances of O, F, Na, Mg and Al.
\label{fig5}}
\end{figure}

\end{document}